\documentclass[aip,jcp,reprint,amsmath,showkeys]{revtex4-1}
\usepackage[version=3]{mhchem}
\usepackage{epstopdf}
\usepackage{siunitx}
\usepackage{dcolumn}
\usepackage[caption=false,labelformat=empty]{subfig}
\usepackage{graphicx}

\draft 

\begin{document}

\title{Evaluation of Constant Potential Method in Simulating Electric Double-Layer Capacitors}

\author{Zhenxing Wang}
\affiliation{Department of Chemistry, University of Kansas, Lawrence, KS 66045, USA}
\author{Yang Yang}
\affiliation{Department of Materials Science and Engineering, University of California, Berkeley, CA 94720, USA}
\author{David L. Olmsted}
\affiliation{Department of Materials Science and Engineering, University of California, Berkeley, CA 94720, USA}
\author{Mark Asta}
\affiliation
{Department of Materials Science and Engineering, University of California, Berkeley, CA 94720, USA}
\author{Brian B. Laird}
\email{blaird@ku.edu}
\thanks{Author to whom correspondence should be addressed}
\affiliation{Department of Chemistry, University of Kansas, Lawrence, KS 66045, USA}

\date{\today}

\begin{abstract}
A major challenge in the molecular simulation of electric double layer
capacitors (EDLCs) is the choice of an appropriate model for the electrode.
Typically, in such simulations the electrode surface is modeled using a uniform
fixed charge on each of the electrode atoms, which ignores the electrode
response to local charge fluctuations in the electrolyte solution. In this
work, we evaluate and compare this Fixed Charge Method (FCM) with the more
realistic Constant Potential Method (CPM), [Reed, et al., J. Chem. Phys.,
\textbf{126}, 084704 (2007)], in which the electrode charges fluctuate in order
to maintain constant electric potential in each electrode. For this comparison,
we utilize a simplified \ce{LiClO4}-acetonitrile/graphite EDLC. At low
potential difference ($\Delta \Psi \le$\SI{2}{V}), the two methods yield
essentially identical results for ion and solvent density profiles; however,
significant differences appear at higher $\Delta \Psi$. At $\Delta \Psi \ge
\SI{4}{V}$, the CPM ion density profiles show significant enhancement (over
FCM) of ``inner-sphere adsorbed'' \ce{Li+} ions very close to the
electrode surface. The ability of the CPM electrode to respond to local charge
fluctuations in the electrolyte is seen to significantly lower the energy (and
barrier) for the approach of \ce{Li+} ions to the electrode surface. 
\end{abstract}


\maketitle

\section{Introduction}
Electric double-layer capacitors (EDLCs) are non-Faradaic, high power-density
devices that have wide application in energy storage. Together with
pseudo-capacitors, EDLCs make up a class of energy storage devices called super
capacitors. The energy storage and release mechanism in EDLCs is rapid and
possesses a long cycle life due to the physical nature of the
charging/discharging process -  in the charging process, ions in the
electrolyte solution aggregate at the interface to form an electric double
layer, which in turn induces charges on the electrode surfaces. Over the past
decade, accompanying the recent bloom of novel electrode (i.e. nanoporous
material) and electrolyte materials (i.e. ionic liquid), EDLCs have been the
subject of numerous experimental studies.\cite{Choi2012,Aravindan2014} From
these studies, the performance of EDLCs is affected by a number of different
factors including, but not limited to, electrolyte
composition\cite{Sato2004,Yuyama2006}, interface structure\cite{Wei2014,Ji2014}
and surface area.\cite{Largeot2008,Itoi2011} As a result of these efforts, the
performance of EDLCs have been significantly enhanced, extending their
applicability. 

These experimental studies have been complemented by a number of
theoretical/computational studies ranging from atomistic simulation to
mesoscale continuum modeling.\cite{Fedorov2014} Analytical continuum models
have been developed to describe the electrode/electrolyte interface, for
example, the Gouy-Chapman-Stern model.\cite{Wang2012,Wang2013} Classical
Density Functional Theories (cDFT)\cite{Jiang2013} have also been applied to
estimate properties of EDLCs, which are more accurate than continuum models
but considerably less computationally demanding than atomistic simulation.
Atomistic simulations, however, have an advantage over the continuum models and
cDFT because they provide a molecular-level description of the structure,
dynamics and thermodynamics of EDLCs. Molecular simulation tools, such as {\em
ab-initio} molecular dynamics (AIMD) simulation \cite{Ando2013} and
classical molecular-dynamics (MD) simulation are widely used to investigate
EDLC interfacial phenomena at the molecular level.

In molecular simulations of EDLCs, the modeling of the electrode is a
particular challenge because of the difficulty in defining a consistent
classical atomistic model for a conductor. In many MD studies of EDLCs, the
electrode atoms are assumed to carry a uniform fixed charge.  This Fixed Charge
Method (FCM), however, neglects charge fluctuations on the electrode induced by
local density fluctuations in the electrolyte
solution.\cite{Yang2009,Yang2010,Shim2012,Feng2011,Feng2013} To explicitly take
into account such fluctuations, the Constant Potential Method (CPM) was
developed by Reed, et al.\cite{Reed2007} This method is based on earlier work
of Siepmann and Sprik \cite{Siepmann1995} in which the constraint of constant
electrode potential was enforced on average using an extended Hamiltonian
approach (similar to the Nos\'e method\cite{Nose1984} for constant
    temperature); however, in the CPM the constraint is applied instantaneously
at every step. Additional corrections to the CPM were added later by Gingrich
and Wilson.\cite{Gingrich2010cpl} In the CPM, the electric potential $\Psi_{i}$
on each electrode atom is constrained at each simulation step to be equal to a
preset applied external potential $V$, which is constant over a given
electrode. This constraint leads to the following equation for the charge,
  $q_{i}$, on each electrode atom (where $i$ indexes the atoms in the
      electrode): 

\begin{equation}
  V=\Psi_{i}=\frac{\partial U}{\partial q_{i}}
  \label{eqn:dudq2v}
\end{equation}
where $U$ is the total Coulomb energy of the system.

The structure of the Coulomb energy expression is such that
Eq.~\ref{eqn:dudq2v} is a system of linear equations for $q_{i}$ and can be
solved with standard linear algebra techniques. To guarantee that the linear
system corresponding to Eq.~\ref{eqn:dudq2v} has a solution, the electrode
point charges are generally replaced with a narrow Gaussian charge
distribution. For a detailed study of the optimal choice for Gaussian width,
see Gingrich\cite{Gingrich2010}. 

Several studies of EDLCs employing the CPM have been
reported. Merlet, et al.\cite{Merlet2011,Merlet2012,Merlet2012jpcc,Merlet2013}
studied a nanoporous carbon electrode in contact with electrolyte consisting of
an ionic liquid or an ionic liquid/acetonitrile mixture. Vatamanu and
coworkers\cite{Vatamanu2010,Vatamanu2012,Vatamanu2012a,Vatamanu2013}
investigated ionic liquid electrolytes with carbon or gold electrodes using
the smooth particle mesh Ewald (SMPE)\cite{Kawata2001,Kawata2002} method to
simplify the calculation. The hydration of metal-electrode surfaces was
examined by Limmer, et al.\cite{Limmer2013prl,Limmer2013pnas} 

In all of these studies, detailed comparisons of the CPM with the fixed charge
method have been lacking. In a recent paper, Merlet, et al.
\cite{Merlet2013jpcl} examined the differences between CPM and FCM simulations
as measured by the relaxation kinetics in EDLC with nanoporous carbide-derived
carbon electrode and the electrolyte structure at interface in EDLC with planar
graphite electrode.It was showed that CPM predicts more reasonable relaxation
time than FCM. For the electrolyte structure, this study showed that there are
some quantitative differences between the results of the two methods, but the
qualitative features were unchanged for these ionic-liquid based EDLCs.

In this work, we study an organic electrolyte/salt-based ELDC,
namely an \ce{LiClO4}/acetonitrile electrolyte at a graphite electrode. To
compare the results for this system using the CPM and FCM, several structural
aspects normally reported in EDLC simulations are studied for comparison,
specifically, the particle and charge density profiles near the electrodes and
the solvation structure of the cation (\ce{Li+}) both in bulk and near the
surface.

\section{Models and Methodology}

\begin{figure}[h]
  \includegraphics[width=0.48\textwidth]{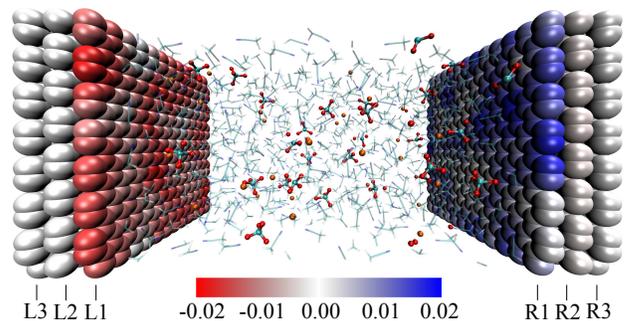}
  \caption{Simulation snapshot at 298K with $\Delta\Psi = 2 $V: negative
    electrode is on left and positive is on right. The color of each electrode
    atom indicates its charge (refer to color scale bar, unit \textit{e}). The
    electrolyte solution is shown between the two electrodes. Orange spheres:
    \ce{Li+}, Red spheres: O in \ce{ClO4-}, cyan spheres: Cl in \ce{ClO4-},
    transparent stick models: acetonitrile. For clarity, in this figure, the
      distance between L1 and R1 (\SI{5.43}{nm}) is smaller than that used in
      the production runs.}
  \label{fig:cell}
\end{figure}

In our simulations, the atoms of the electrolyte solution are placed between
two carbon electrodes, each consisting of three graphite layers. The simulation
geometry is shown in Fig.~\ref{fig:cell}, which shows a snapshot from a
simulation at 298K with a potential difference, $\Delta \Psi$ of \SI{2}{V}. For
the production runs, the distance between the two inner-most electrode layers
(labeled L1 and R1, respectively, in Fig.~\ref{fig:cell}) is {\SI{6.365}{nm}},
which is far beyond the Debye length (\SI{0.2}{nm}). The electrolyte between
the electrodes consists of 588 acetonitrile molecules and 32
\ce{Li+}/\ce{ClO4-} pairs, corresponding to a LiClO$_4$ concentration of
\SI{1.00}{M}. The total dimensions of the cell are 2.951, 2.982 and
\SI{8.040}{nm}. The positions of the electrode carbon atoms are held fixed
during the simulation. 

To model the molecular interactions we employ a variety of literature force
fields. For acetonitrile, we use the united atom model of Edwards, et
al.\cite{Edwards1984} For the ions, we use the forcefield of Eilmes and
Kubisiak,\cite{Eilmes2011} excluding the polarizability terms. The interaction
parameters for the graphite electrode carbon atoms are taken from
Ref.~\onlinecite{Cole1983}. Lorentz-Berthelot mixing rules are used to
construct all cross interactions. 

In this slab geometry, we define the \textit{z}-axis as the direction normal to the
electrodes and apply periodic boundary conditions  only in the {\em x-y} plane
(parallel to the graphite layers).  Unlike the original CPM, which used
2d-periodic Ewald sums,\cite{DeLeeuw1979,Kawata2001} 3d-periodic Ewald sums
with shape corrections\cite{Yeh1999} were used in this work to improve the
calculation speed, with a volume factor set to 3. The correction term to the
usual 3d-periodic energy expression is given by (see Eq. A.21 in
Ref.~\onlinecite{Gingrich2010}):
\begin{equation}
  \frac{2\pi}{V}\left(\sum_{i}q_{i}z_{i}\right)^{2}
\end{equation}

In studies using FCM, either the uniform charge on each electrode atom is either
arbitrarily chosen\cite{Yang2010,Shim2012,Feng2013} or is estimated using a
time-consuming trial and error procedure to yield the
specified electric potential difference,\cite{Feng2011} which is calculated by
numerically integrating the Poisson equation using the charge density profile.
The CPM is able to predict the explicit average charge on each electrode atom at a
given potential difference; therefore, for consistency, in our FCM simulations
we set the charge on each electrode atom to the average charge per atom
obtained by the CPM calculations at the same potential difference. Otherwise,
all other force-field parameters in the FCM simulations are identical to those
used in the CPM simulations.

All simulations were performed using the molecular-dynamics simulation code
LAMMPS,\cite{Plimpton1995} modified to implement the CPM, using a time step of
{\SI{1}{fs}}.  Constant $NVT$ conditions are enforced using a Nos\'{e}-Hoover
thermostat with a relaxation time of 100 fs and a temperature of \SI{298}{K}. The
cutoffs for all non-bonded interactions are {\SI{1.4}{nm}}. For the Ewald sums,
an accuracy (relative RMS error in per-atom forces) is set to $10^{-8}$. The
parameter of the Gaussian electrode charge distribution (see Eq. S2 in
supplementary material) is set to {\SI{19.79}{nm^{-1}}, which is the same as in
  Ref.~\onlinecite{Reed2007}.  All results reported here are statistical
    averages taken from runs of 25 to {\SI{30}{ns}} in length, each preceded by
    {\SI{2}{ns}} of equilibration.  Further details as to the CPM method and
  our implementation in LAMMPS can be found in the supplementary material.

\section{Results and discussion}

\subsection{Electrode charge distribution}
Unlike the FCM, the CPM allows the individual atom charges on the electrode to
fluctuate in response to local charge rearrangements in the electrolyte. The
total charges on each layer for several potential differences are shown in
Table~\ref{tab:layercharge}. The net electrode charge for each simulation does
not show values statistically significant from zero, as expected. In a perfect
conductor, the charges on the electrode atoms would be concentrated entirely on
the electrode surface layers (L1 and R1 in Fig.~\ref{fig:cell}); however, in
the CPM the charge distribution on the electrode is approximated by discrete
point charges centered on the electrode itself, so a small amount of charge is
found on the second layer and to a much lesser extent the third.

\begin{table}
\caption{Average total charge on each electrode layer\protect\footnote{The numbers in
  parentheses represent 95\% confidence intervals in the last significant
    figure}}
\label{tab:layercharge}
\begin{ruledtabular}
\begin{tabular}{cdddd}
Layer & \multicolumn{1}{c}{\SI{0}{V}} & \multicolumn{1}{c}{\SI{2}{V}} & \multicolumn{1}{c}{\SI{4}{V}} & \multicolumn{1}{c}{\SI{5}{V}}\\
\hline
L3     & 0.00(2)    & 0.03(2)    & 0.06(2)    & 0.06(2)   \\   
L2     & 0.02(5)    & 0.33(5)    & 0.64(4)    & 0.75(5)   \\
L1     &-0.1(4)     &-3.4(4)     &-6.6(4)     &-8.4(5)    \\
R1     & 0.0(4)     & 3.3(4)     & 6.5(4)     & 8.4(5)    \\
R2     & 0.01(5)    &-0.29(4)    &-0.58(5)    &-0.75(6)   \\
R3     & 0.00(2)    &-0.02(2)    &-0.03(2)    &-0.06(2)   \\
Total & 0.01(6)    & 0.02(6)    & 0.04(6)    & 0.01(6)   \\
\end{tabular}
\end{ruledtabular}
\end{table}

Figure~\ref{fig:chargehis} is a log-linear plot of the probability distribution
of individual electrode atom charges, $p(q)$, on the inner electrode layers for
various potential differences. For the inner layer of the positively charged
electrode (R1), the charge distribution is well described by a Gaussian
distribution over the entire range of potential differences studied
($\Delta\Psi$ = \SI{0}{V} to \SI{5}{V}). For the negatively charged electrode (L1), this is
true at the lower potential differences (up to $\Delta\Psi$ = \SI{2}{V}), but the
charge distribution takes on a bimodal structure at higher potential
differences ($\Delta\Psi$ = \SI{4}{V} and \SI{5}{V}). A similar non-Gaussian charge
distribution has also been seen in simulations at the negative electrode of a
\ce{H2O}/Pt system;\cite{Limmer2013prl} however, that system differs somewhat
from our simulated system in that no ions are present. The bimodal charge
distribution at the higher potential differences in this system highlights an
important difference between the CPM and the FCM. Unlike the FCM, the electrode
charges in the CPM can adjust to respond to local fluctuations in the
electrolyte/ion charge density. As we will see in the next section, this second
peak in $p(q)$ seen at high potential differences is due to the presence of
\ce{Li+} ions near the electrode surface, which induce higher than average local
charges on the adjacent electrode. 

\begin{figure}[h]
  \includegraphics[width=0.5\textwidth]{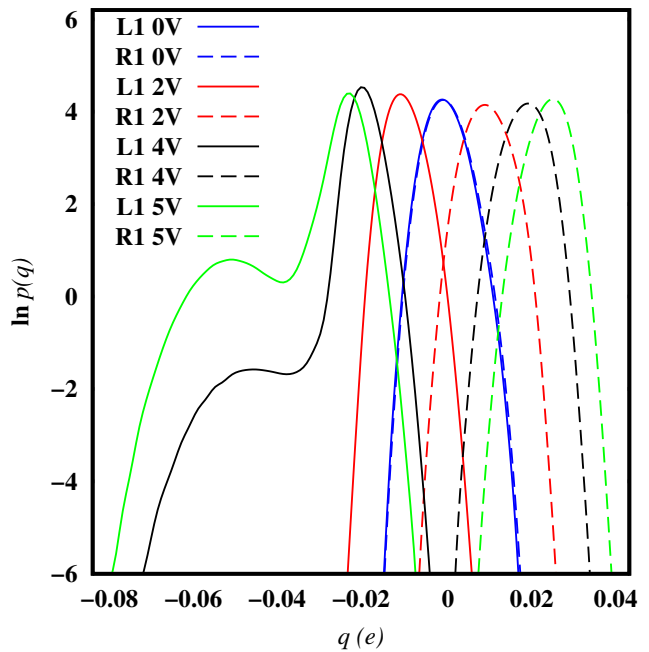}
  \caption{Distribution of electrode atom charges for the inner electrode
  layers at various potential differences: L1 (solid lines) and R1(dashed
  lines).}
  \label{fig:chargehis}
\end{figure}

\subsection{Density profiles for ions and electrolytes}
To better understand the origin of the bimodal charge distributions in the CPM
at high potential difference, we examine the ion density profiles at the
interface and compare the results obtained using CPM and FCM
(Fig.~\ref{fig:iondp}). 
\begin{figure}[h]
  \includegraphics[width=0.5\textwidth]{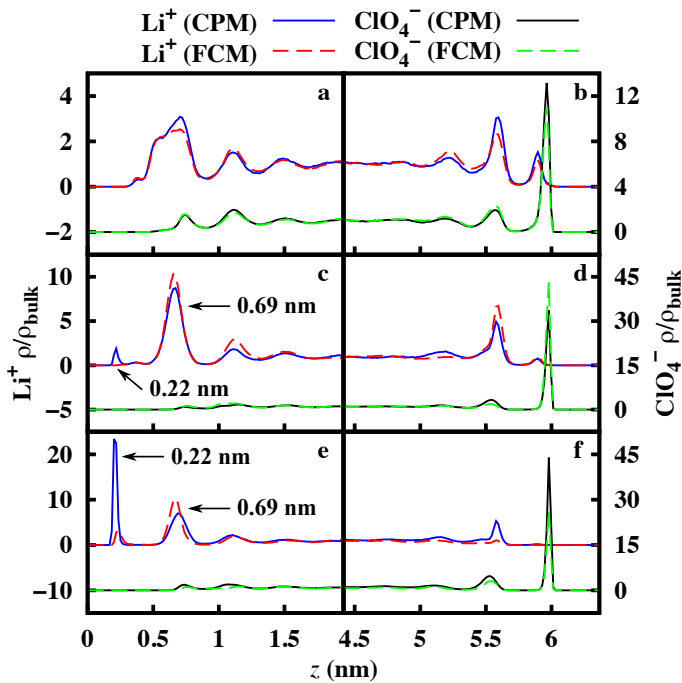}
  \caption{Ion densities near each of the two electrodes (L1 and R1) for
    various applied potential differences (a) L1; $\Delta\Psi= \SI{2}{V}$, (b)
      R1; $\Delta\Psi=\SI{2}{V}$, (c) L1; $\Delta\Psi=\SI{4}{V}$ (d) R1 $\Delta\Psi=
      \SI{4}{V}$ (e) L1; $\Delta\Psi= \SI{5}{V}$, (f) R1; $\Delta\Psi=
      \SI{5}{V}$. The negatively and positively charged electrodes (L1 and R1)
      are at $z = 0$ and \SI{6.365}{nm}, respectively.}
  \label{fig:iondp}
\end{figure}

At the lowest non-zero potential difference ($\Delta\Psi$ = \SI{2}{V}), the ion
density profiles for the CPM and FCM methods show relatively minor quantitative
differences in peak heights, but otherwise the density peak positions and
overall structure are identical. At higher potential differences (\SI{4}{V} and \SI{5}{V}),
however, qualitative differences emerge. At $\Delta\Psi$ = \SI{4}{V}, a peak in the
\ce{Li+} density profile near the negatively charged electrode (L1) at
\SI{0.22}{nm} appears in the CPM calculation, but is absent in the FCM
results. This peak represents inner-sphere adsorbed \ce{Li+} ions that no
longer possess a full solvation shell of acetonitrile, but are also ``partially
solvated'' by the electrode. The existence of these inner-sphere adsorbed ions
is made possible because the CPM allows for more physically correct electrode
charge distribution - one that can respond to the presence of the nearby
\ce{Li+} ion with locally larger-than-average negative electrode charges. The
partial desolvation was reported to be highly related with the confinement of
ions, which fundamentally affects the capacitance.\cite{Merlet2013nc}

As the potential difference is increased beyond \SI{4}{V}, the inner-sphere
adsorbed ion peak grows substantially. For the FCM, we see this peak at
\SI{5}{V}, but it has an amplitude that is much smaller than that predicted by
the CPM. In addition, Fig.~\ref{fig:iondp} also shows that the height of the
\ce{Li+} density peak at \SI{0.69}{nm}, representing the first fully solvated
outer-sphere adsorbed \ce{Li+} layer in the ELDC, is overestimated at $\Delta
\Psi =$ \SI{4}{V} and \SI{5}{V} in the FCM. This is due to the migration of
\ce{Li+} ion density from the inner-spher adsorbed peak at \SI{0.21}{nm},
compared with the CPM.

In contrast to \ce{Li+}, the \ce{ClO4-} density profiles are nearly identical
for both methods at all studied potential differences. The charge density in
\ce{ClO4-} is spread out over a larger radius and does not create as large a
local charge concentration in the electrolyte solution to perturb the electrode
charge distribution enough to affect the density profiles. 

\begin{figure}[h]
  \includegraphics[width=0.5\textwidth]{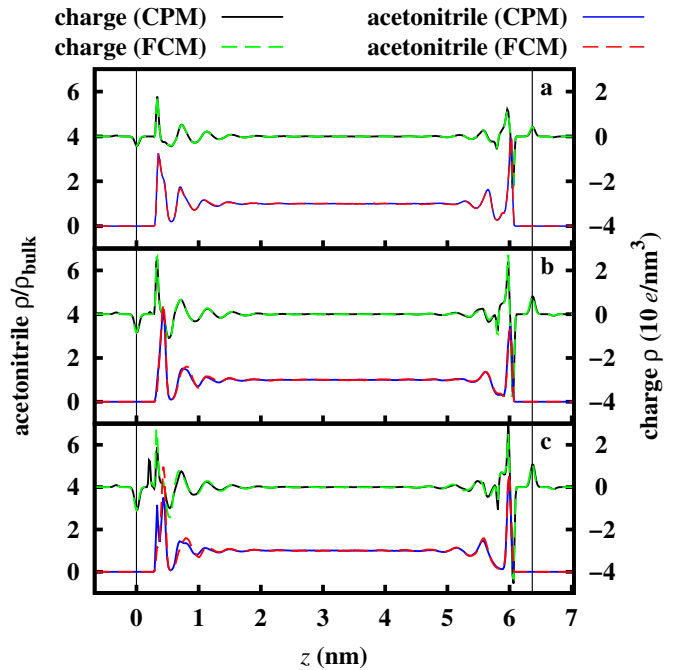}
  \caption{Acetonitirle density and charge profiles at various applied
    potentials (a) $\Delta\Psi=\SI{2}{V}$, (b) $\Delta\Psi=\SI{4}{V}$, (c)
      $\Delta\Psi=\SI{5}{V}$. Vertical lines show the position of the
      electrode inner layers L1 (left) and R1 (right).}
  \label{fig:acndp}
\end{figure}

In Fig.~\ref{fig:acndp} we plot the solvent (acetonitrile, center of mass) and
charge density profiles.  Unlike the ion density profiles, the density profiles
for acetonitrile are identical for both methods up to a potential difference of
4V. However, at $\Delta\Psi =$ 5V the acetonitrile peak closest to the negative
electrode (L1) splits into two peaks in the CPM, a feature not seen in the FCM
calculation. This is because the large amount of electrode solvated \ce{Li+}
seen in the CPM pulls the acetonitrile molecules in its solvation shell closer
to the surface. 

The structure of the electric double layer in this system is best illustrated
by the charge density profile (see Figure~\ref{fig:acndp}). For this quantity
the results for both methods (CPM and FCM) are very similar with the only
difference coming at high potential difference, where small charge peaks
corresponding to the electrode-solvated \ce{Li+} are present in the CPM result,
but not in the FCM.  

\subsection{Structure of inner-sphere adsorbed \ce{Li+}}

\begin{figure}[h]
  \subfloat{\includegraphics[width=0.5\textwidth]{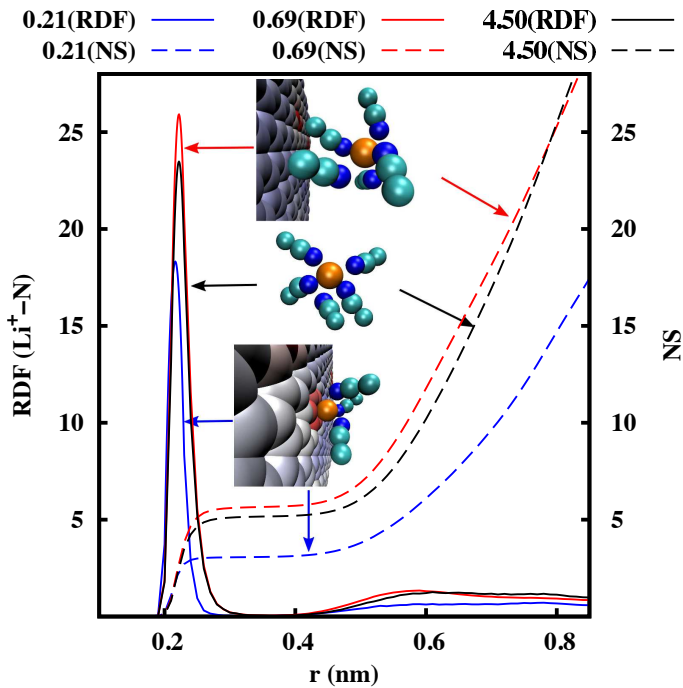}}
  \caption{Radial distribution functions (RDF,solid lines) and corresponding
  number of solvent molecules (NS, dashed lines) between \ce{Li+} and N atom in
  acetonitrile for $\Delta\Psi$ = \SI{4}{V}. In the legend the number
  indicates the distance (nm) from the negative electrode surface to the center
  of bin (with width {$\pm$\SI{0.02}{nm}}) containing the \ce{Li+} ion. The
  insets show typical snapshots of the \ce{Li+} solvation structure for each
  distance studied.} \label{fig:rdfns}
\end{figure}

In this work, the principal difference between the results derived from the CPM
and FCM is the appearance in the CPM of \ce{Li+} ions very close to the
negatively charged electrode at high potentials.  This inner-sphere adsorbed
\ce{Li+} also is seen in the FCM, but only at very low concentration at the
highest potential difference studied. To examine this feature more closely, we
analyze the solvation structure of \ce{Li+} by plotting in Fig.~\ref{fig:rdfns}
(for $\Delta\Psi = \SI{4}{V}$) the CPM radial distribution function (RDF)
between the \ce{Li+} ion and the acetonitrile N for three \ce{Li+} distances
from the electrode surface: 0.21, 0.69 and \SI{4.50}{nm}. These correspond to
the inner-sphere adsorbed \ce{Li+}, the first layer of outer-sphere adsorbed
\ce{Li+} and the bulk conditions, respectively. Also plotted in
Fig.~\ref{fig:rdfns} are the number of solvent molecules (NS) within a given
radius - for the bulk ions, the first plateau in this quantity corresponds to
the equilibrium coordination number. The results of RDF and NS for the FCM or
other potential differences are identical to that shown for the CPM at
\SI{4}{V} (Fig.~\ref{fig:rdfns}) - except for the fact that no \ce{Li+} ions at
\SI{0.21}{nm} for the CPM at $\Delta\Psi = \SI{2}{V}$ and for the FCM at
\SI{2}{V} and \SI{4}{V} are detected and therefore no RDF (nor NS) data was
obtained. 

At $z$ = \SI{0.69}{nm} and \SI{4.50}{nm}, the \ce{Li+} ion is fully solvated by the
acetonitrile solvent and the first solvation shell peaks for both distances are
nearly identical with a peak distance of \SI{0.22}{nm}. In addition, the coordination
number (given by the plateau value of NS) of each \ce{Li+} ion is equal to
about 5.0 for the ``bulk'' case (\SI{4.50}{nm}), just slightly smaller than that for the
first fully solvated peak with 5.4, presumably due to the enhanced acetonitrile
density near the electrode region (see Fig.~\ref{fig:acndp}). The value for the
``bulk'' is consistent with that seen in other simulation studies of bulk
acetonitrile-lithium salt solutions.\cite{OliveiraCosta2010,Spangberg2004a} 

For \ce{Li+} ions at \SI{0.21}{nm} from the electrode, however, the solvent
coordination is significantly reduced from the bulk value to 3.1 due to
partial solvation by the electrode atoms themselves. The bottom inset in
Fig.~\ref{fig:rdfns} shows a representative snapshot from the CPM simulation
($\Delta\Psi = \SI{4}{V}$) of an electrode-solvated ion and its nearby
environment. In the CPM, because the charges on the electrode can fluctuate
individually in response to the local electrolyte charge distribution, the
presence of the \ce{Li+} ion very close to the electrode induces larger than
average negative charges on the nearby electrode ions, as seen in the inset.
The ability of the electrode to respond to local fluctuations is necessary for
the stabilization of the electrode-solvated \ce{Li+} below $\Delta\Psi \lesssim
\SI{4}{V}$, as no such ions are seen in the FCM simulations in this range. At
\SI{5}{V}, such electrode-solvated ions are seen in the FCM, but at
significantly lower concentration than in the CPM, indicating a much higher
energy for such configurations in the FCM, relative to the CPM. We have
confirmed that sampling in the simulations is sufficient, as we see multiple
crossings and recrossings of \ce{Li+} ions between the positions at
\SI{0.69}{nm} and \SI{0.21}{nm}.

\section{Summary}
In this work, two methods, the Constant Potential Method (CPM) and the Fixed
Charge Method (FCM), are compared in a simulation of a model for a
\ce{LiClO4}-acetonitrile/graphite electric double-layer capacitor. The major
difference between the two methods is that in the CPM the charges on the
individual electrode atoms can fluctuate in response to local fluctuations in
electrolyte charge density, whereas those charges for the FCM are static. For
this system, there are no measurable differences between the results of these
two methods at low potential differences ($\Delta\Psi \le \SI{2}{V}$); however,
at larger potential differences significant qualitative differences emerge in
the EDLC ion spatial distribution. 

At a potential difference of 4V, a new peak in the \ce{Li+} density profile
appears in the CPM calculation at 0.21 nm away from the electrode. This peak -
absent in the FCM calculation - corresponds to a inner-sphere adsorbed \ce{Li+}
ion that is close enough to the electrode to have lost some of its acetonitrile
solvation shell and is partially solvated by the electrode atoms. For this
electrode-solvated \ce{Li+} ion the acetonitrile coordination number is reduced
from about 5.0 for a fully solvated ion to 3.1. This partial solvation is
possible because in the CPM the electrode atom charges can respond to local
fluctuations in the electrolyte charge density  - in this case negative charges
build up in the CPM electrode near the lithium ion. This ability lowers the
energy of the ``electrode-solvated'' \ce{Li+} ion relative to that of a
fixed-charge electrode (as in the FCM). This close approach of \ce{Li+} to the
electrode is possible in the FCM, but only at higher electrode potential
differences and a considerably lower concentrations than when CPM is used. The
energetics of the approach of a \ce{Li+} ion to the electrode is important in
many pseudo-capacitor applications, and, as our calculations show, the CPM
would be preferable to the FCM in the calculation of the barrier energy of this
process because it more accurately represents the fluctuating charges on the
electrode. 

\begin{acknowledgements}

The authors thank Dr. David Limmer for helpful discussions on the constant
potential method. This work is supported as part of the Molecularly Engineered
Energy Materials, an Energy Frontier Research Center funded by the US
Department of Energy, Office of Science, Office of Basic Energy Sciences under
Award No.  DE-SC0001342.

\end{acknowledgements}

\pagebreak
\widetext
\setcounter{equation}{0}
\setcounter{figure}{0}
\setcounter{table}{0}
\setcounter{page}{1}
\makeatletter
\renewcommand{\theequation}{S\arabic{equation}}
\renewcommand{\thefigure}{S\arabic{figure}}

\begin{center}
\textbf{\large Supplemental Materials: Evaluation of Constant Potential Method in Simulating
 Electric Double-Layer Capacitors}\\
\vspace{0.4cm}
Zhenxing Wang$^1$ Yang Yang$^2$, David L. Olmsted$^2$, Mark Asta$^2$ and Brian B. Laird$^{*1}$\\
\vspace{0.2cm}
$^1${\em  Department of Chemistry, University of Kansas, Lawrence, KS 66045, USA}\\
$^2$ {\em Department of Materials Science and Engineering, University of California, Berkeley, CA 94720, USA}\\
\end{center}
\noindent
$^*$ Author to whom correspondence should be addressed

\vspace{0.4cm}

\noindent{\bf Table of Contents}
\begin{itemize}
\item[{\bf S1}] Details of Constant Potential Method
\item[{\bf S2}] Implementation of CPM in LAMMPS
\end{itemize}

\vskip 0.6cm
\noindent
{\bf S1 {\em Constant Potential Method}}\\
In what follows, the charge and position vector of an electrode (electrolyte) atom
are designated as $Q_{i}$ ($q_{j}$) and $\textbf{R}_{i}$ ($\textbf{r}_{j}$),
respectively. Starting from Eq.\,1 in the main text, the potential at the position
of a selected electrode atom ($l$) with the charge as $Q_{l}$ and the positon as
$\textbf{R}_{l}$ can be calculated by:
\begin{equation}
\Psi_{l}=\Psi_{l}^\textrm{kspace}+\Psi_{l}^\textrm{real}+
  \Psi_{l}^\textrm{self}+\Psi_{l}^\textrm{slab}
\end{equation}

Here $\Psi_{l}^\textrm{kspace}$ the $k$-space contribution from the 3-d Ewald
sum; $\Psi_{l}^\textrm{real}$ is the corresponding real-space contribution;
$\Psi_{l}^\textrm{self}$ is the self-correction and $\Psi_{l}^\textrm{slab}$ is
the correction for the slab geometry. To guarantee the solvability of the CPM
linear equations, Gaussian charges are used for electrode atoms\cite{Gingrich2010}
\begin{equation}
Q_{i}(\textbf{r})=Q_{i}\left(\frac{\eta^2}{\pi}\right)^{3/2}e^{-\eta^2\left(\textbf{r}-\textbf{R}_i\right)^2}
\end{equation}

Here $\eta$ is the parameter defining the width of the distribution.

Each portion of the potential for the mixed electrode (Gaussian charge) and
electrolyte (point charge) system are obtained as follows.

\vspace{0.2cm}
\textbf{Real-space contributions}: The Ewald expression for the real-space
contribution is
\begin{equation}
\Psi_{l}^\textrm{real}=
  \frac{1}{2}\sum_{i=1}^{n^\prime}Q_{i}\frac{\textrm{erfc}\left({\alpha}\,r_{li}\right)
    -\textrm{erfc}\left(\frac{\eta}{\sqrt{2}}r_{li}\right)}{r_{li}}+
  \sum_{j=1}^{m}q_{j}\frac{\textrm{erfc}\left({\alpha}\,r_{lj}\right)
  -\textrm{erfc}\left({\eta}\,r_{lj}\right)}{r_{lj}}
\end{equation}
Where $n^\prime$ denotes the $i=l$ term is excluded.
$r_{li}=\lvert\textbf{R}_{l}-\textbf{R}_{i}\rvert$ and
$r_{lj}=\lvert\textbf{R}_{l}-\textbf{r}_{j}\rvert$.

\vspace{0.2cm}
\textbf{{\em k}-space contributions}: The Ewald expression for the $k$-space
contribution is 
\begin{equation}
\Psi_{l}^\textrm{kspace}=\frac{1}{V}\sum_{\textbf{k}>0}\Gamma(\kappa)
  \left[e^{-i\,\textbf{k}{\cdot}\textbf{R}_{l}}S(\textbf{k})+
  e^{i\,\textbf{k}{\cdot}\textbf{R}_{l}}S(-\textbf{k})\right] \label{keq}
\end{equation}

Following the notation in Refs.~\onlinecite{Moore,Gingrich2010}, here
$\textbf{k}$ is a reciprocal lattice vector, $\kappa=\lvert{\textbf{k}}\rvert$
and $\Gamma(\kappa)$ is the Fourier coefficient of the Gaussian function used
in the Ewald sum 
\begin{equation}
\Gamma(\kappa)=\frac{4{\pi}e^{-\kappa^2/4\alpha^2}}{\kappa^{2}}
\end{equation}

In Eq. \ref{keq}, 
$S(\textbf{k})$ is the structure factor, given as
\begin{equation}
S(\textbf{k})=\sum_{i=1}^{n}Q_{i}e^{i\,\textbf{k}{\cdot}\textbf{R}_{i}}+
  \sum_{j=1}^{m}q_{j}e^{i\,\textbf{k}{\cdot}\textbf{r}_{j}}
\end{equation}

\vspace{0.2cm}
\textbf{Self correction}: 
\begin{equation}
\Psi_{l}^\textrm{self}=\frac{\sqrt{2}\eta-2\alpha}{\sqrt{\pi}}Q_{l}
\end{equation}

\vspace{0.2cm}
\textbf{Slab correction}:
\begin{equation}
\Psi_{l}^\textrm{slab}=\frac{4\pi Z_{l}}{V}\left(\sum_{i=1}^{n}Q_{i}Z_{i}+\sum_{j=1}^{m}q_{j}z_{j}\right)
\end{equation}
where $Z_{i}$ is the $z$ component electrode atom position
and $z_{j}$ is the corresponding quantity for an electrolyte atom.

The potential expression can then be rewritten as
\begin{equation}
\Psi_{l}=\sum_{i}^{n}a_{li}Q_{i}+\sum_{j}^{m}b_{lj}q_{j}
\end{equation}
where $a_{li}$ is a function of the position of selected electrode atom
($\textbf{R}_{l}$) and that of other electrode atoms ($\textbf{R}_{i}$) only,
which are constant because all electrode atoms are fixed. In opposite,
$\textbf{r}_{j}$ is varying during the simulation and thus ${b_{lj}}$ is
updated every timestep. Because $\Psi_{l}$ is equal to the external potential
$V$, above equation becomes

\begin{equation}
\sum_{i}^{n}a_{li}Q_{i}=V-\sum_{j}^{m}b_{lj}q_{j}
\end{equation}

Similarly we can obtain a linear equation for each electrode atom. With
$\textbf{Q}=[Q_{1},Q_{2},\cdots,Q_{n}]$ as unknown, the linear system can be
then written as \begin{equation} \textbf{A}\textbf{Q}=\textbf{V}-\textbf{B}
\end{equation}

By solving this well-determined linear equations system the explicit charge for
electrode atoms can be determined.

\vskip 0.6cm
\noindent
{\bf S2 {\em Implementation in LAMMPS}}

The implemented code includes three functions: setting up the linear system of 
equations, solving the linear system and updating atom charges, energy and
per-atom forces. The code can be downloaded through https://code.google.com/p/lammps-conp/

Determining the real-space contribution to $\textbf{A}$ and $\textbf{B}$ is
similar to the real-space Coulombic energy calculation in LAMMPS, but without
using the table technique (see the \textit{table} keyword in pair\_modify command
in the LAMMPS manual for details). The self correction and slab correction
portion are straightforward and the approach is similar to that of the
corresponding components of the energy calculation in LAMMPS. However the Ewald sum
contribution is obtained in a different way. Elements in $\textbf{A}$ depend on
the positions of electrode atoms only, which is constant during the simulation
(assuming fixed electrode atoms). Therefore, at the initial stage, all MPI
tasks collect the coordinates of all electrode atoms and calculate Ewald sum
contribution to $\textbf{A}$ individually. In contrast, $\textbf{B}$ is only
the summation of the contribution from only electrolyte atoms, so that Ewald sum
portion in $\textbf{B}$ is calculated in a similar way as the Ewald sum in
group/group interaction calculation in LAMMPS, which each MPI task only
calculates the portion in its atom list and then sum them together.

At the initial stage, the matrix inverse of $\textbf{A}$ is obtained using
LAPACK. In each step, at the second half step of the velocity-Verlet, the
$\textbf{B}$ is calculated and the charges of electrode atoms obtained, after
which the additional energy and per-atom force are calculated (See
reference~\onlinecite{Gingrich2010} for details of the additional terms) and
added to original energy and per-atom force.

\end{document}